\documentclass[%
 reprint,
superscriptaddress,
amsmath,amssymb,
aps]{revtex4-2}

\usepackage{graphicx}
\usepackage{float}
\usepackage{dcolumn}
\usepackage{bm}
\usepackage{physics}
\usepackage{xcolor}
\usepackage{siunitx}

\definecolor{gio}{rgb}{0.82, 0.34,0.80}
\definecolor{NiCCo}{rgb}{0.91,0.48,0.2}
\definecolor{luca}{rgb}{0.4,0.78,0.77}
\definecolor{giu}{rgb}{1,0,0}
\definecolor{nic}{rgb}{0.45,0.38,0.96}

\graphicspath{{./figure/}}

\usepackage{filecontents}

\begin{document}

\preprint{APS/123-QED}

\title{Blue repulsive potential for dysprosium Bose-Einstein condensates}

\author{Niccolò Preti}
\email{niccolo.preti@unifi.it}
\affiliation {Dipartimento di Fisica e Astronomia, Università di Firenze}%
\affiliation {European Laboratory for Nonlinear Spectroscopy (LENS), Università di Firenze}
\affiliation{Consiglio Nazionale delle Ricerche - Istituto Nazionale di Ottica, sede secondaria di Pisa}%

\author{Nicolò Antolini}%
\affiliation {Dipartimento di Fisica e Astronomia, Università di Firenze}
\affiliation {European Laboratory for Nonlinear Spectroscopy (LENS), Università di Firenze}
\affiliation{Consiglio Nazionale delle Ricerche - Istituto Nazionale di Ottica, sede secondaria di Pisa}%

\author{Giulio Biagioni}%
\affiliation {Dipartimento di Fisica e Astronomia, Università di Firenze}
\affiliation {European Laboratory for Nonlinear Spectroscopy (LENS), Università di Firenze}
\affiliation{Consiglio Nazionale delle Ricerche - Istituto Nazionale di Ottica, sede secondaria di Pisa}%

\author{Andrea Fioretti}%
\affiliation {Consiglio Nazionale delle Ricerche - Istituto Nazionale di Ottica, sede secondaria di Pisa}

\author{Giovanni Modugno}%
\affiliation {Dipartimento di Fisica e Astronomia, Università di Firenze}
\affiliation {European Laboratory for Nonlinear Spectroscopy (LENS), Università di Firenze}
\affiliation{Consiglio Nazionale delle Ricerche - Istituto Nazionale di Ottica, sede secondaria di Pisa}%

\author{Luca Tanzi}%
\affiliation {Consiglio Nazionale delle Ricerche - Istituto Nazionale di Ottica, sede secondaria di Pisa}
\affiliation {European Laboratory for Nonlinear Spectroscopy (LENS), Università di Firenze}

\author{Carlo Gabbanini}
\email{carlo.gabbanini@ino.cnr.it}
\affiliation{Consiglio Nazionale delle Ricerche - Istituto Nazionale di Ottica, sede secondaria di Pisa}%

\date{\today}
\begin{abstract}

Short-wavelength repulsive potentials for quantum gases allow to realize new systems and to study new phenomena. Here we report the realization of repulsive optical potentials for dysprosium atoms in the blue region of the spectrum, at wavelengths close to 400 nm. We employ a spectrally-filtered diode laser system to measure both scalar and tensorial components of the polarizability of dysprosium, which we find in good agreement with the theoretical predictions. We demonstrate the {implementation} of potential strengths appropriate to manipulate Bose-Einstein condensates, with scattering-limited lifetimes exceeding one second. This type of optical potentials opens interesting directions for the study of dipolar superfluids and supersolids.
\end{abstract}

\maketitle

\section{\label{sec:intro}Introduction}

Short-wavelength, repulsive optical potentials are key tools to study fundamental phenomena in ultracold quantum gases. Notable examples are uniform Bose \cite{Ga13} and Fermi gases \cite{Mu17}, device-like systems \cite{Kr15}, controllable disorder \cite{Bi08}, and controllable vortexes \cite{Kwo2021}. Repulsive potentials are favored over attractive ones by the fact that the atoms reside mostly in the dark, limiting negative effects such as inelastic light scattering or unwanted disorder due to optical fringes. Since the minimum length scale of an optical potential is set by the related wavelength, through diffraction, it is necessary to use light towards the blue region of the visible spectrum to achieve the highest spatial resolution.

Strongly dipolar lanthanide atoms like dysprosium or erbium are opening new directions of study in the context of quantum gases, because the presence of the long-range dipolar interaction leads to novel phenomena and phases, such as quantum droplets \cite{Ka16}, supersolids \cite{Ta19,Bo19,Ch19} and dipolar solids \cite{Su_2023}. So far, the realization of optical potentials for lanthanides has however been limited to the near-infrared or mid-visible regions of the spectrum \cite{ravensbergen2018accurate,kreyer2021measurement,chalopin2018anisotropic,becher2018anisotropic, Bl23}, despite the fact that atomic species possess strong absorption lines in the blue region. One of the complexities of lanthanides is the relevance of all three components of the polarizability, scalar, vectorial and tensorial, which makes both the theoretical analysis and the experimental control of the optical polarizability more challenging than for other atomic species \cite{dzuba2011dynamic,li2016optical, Bl24}.
This, combined with the multitude of spectral lines present in magnetic lanthanides, has made it difficult to achieve agreement between experiment and theory in spectral regions  close to strong or weaker lines \cite{becher2018anisotropic,Bl24}. The evaluation of the polarizability close to strong transitions can thus be a relevant tool to test both the theory and the atomic spectroscopic parameters.

In this work, we study the realization of repulsive potentials for dysprosium at wavelengths close to the group of strong lines from 405~nm to 421~nm. We investigate experimentally the region between 403 and 404~nm, developing a tunable laser system based on commercially available diode lasers, spectrally filtered to avoid near-resonant components of the background emission. Studying the mechanical effect of light on a dysprosium Bose-Einstein condensate in a combined infrared-blue potential, we measure both scalar and tensorial components of the polarization, finding a good agreement with the theory. We demonstrate that with few milliWatts of light, it is possible to obtain potential depths of the order of hundreds of nanoKelvin, with lifetimes of the order of one second. Our findings, in combination with the theory, also give guidance for realizing optical potentials for dysprosium in a larger region of the blue spectrum, from 400 to 425~nm.

These results open the possibility of realizing new types of experiments for dipolar superfluids and supersolids. For example, an interesting device-like configuration that one could realize is the annular geometry that was proposed a long time ago by A. J. Leggett to test the differences between superfluids and supersolids under rotation \cite{Le70}.
Moreover, the annular geometry makes it possible to realize naturally quasi-1D dipolar superfluids and supersolids with homogeneous density profiles. In contrast, simpler box-like potentials, often employed to study homogeneous non-dipolar systems, exhibit an enhanced inhomogeneity due to the unscreened dipolar interaction at the trap edges. \cite{Roccuzzo22}.
Repulsive optical tweezers could also be used to generate deterministically vortices in dipolar quantum gases as done previously in non-dipolar ones \cite{Kwo2021}, as an alternative to the rotation-induced vortexes demonstrated so far \cite{Klaus2022,casotti2024observation}. In addition, short-wavelength random optical potentials could be employed to study disorder-related localization phenomena \cite{Deissler2010,Schreiber2015} also in dipolar quantum gases, where the long-range nature of the dipole-dipole interaction is expected to produce interesting new effects \cite{Nandkishore2017}.

\section{\label{sec:theory}Theory}

\subsubsection{\label{sec:level3} Atomic polarizability}
When an atom interacts with a non-resonant laser field described by a polarization vector $\bm{e}$ and angular frequency $\omega$, the AC Stark shift of the atomic ground state produces an effective atom-light potential that can be modeled as \cite{Gr00}:
\begin{equation}
    U(\bm{r},\omega)=-\frac{\text{Re}[\alpha(\omega)]}{2\epsilon_0 c}I(\bm{r})\,.
\end{equation}
Here $\epsilon_0$ represents the vacuum permittivity, $c$ the speed of light in vacuum, $I(\bm{r})=\epsilon_0 c\abs{E(\bm{r})}^2/2$ the light intensity and $\alpha(\omega)$ is a quantity characterizing the strength of the atom-light interaction called dynamical polarizability. Generally speaking, the dynamical polarizability can be decomposed into scalar, vectorial, and tensorial contributions. In lanthanides like dysprosium and erbium, because of their open $f$ shell structure, all three components of the polarizability are in general relevant and so the theoretical investigation of the atomic polarizability is more challenging than in other atomic species \cite{dzuba2011dynamic,li2016optical}. It is therefore important to achieve an experimental confirmation of the theory predictions \cite{Kr21,ravensbergen2018accurate,chalopin2018anisotropic}.
When considering a linearly polarized laser field the vectorial contribution, being proportional to $\abs{\bm{e}\cross\bm{e}^*}$, vanishes and the quantity $\alpha(\omega)$ can be written for a state with non-zero angular momentum \cite{Mi10}:
\begin{equation} \label{eq:firstalphaofomega}
    \alpha(\omega)=\alpha_s(\omega)+\frac{3M_J^2-J(J+1)}{J(2J-1)}\frac{3\cos^2\theta-1}{2}\alpha_t(\omega)\,.
\end{equation}
Here $J$ represents the total electronic angular momentum of the atomic ground state and $M_J$ is the projection of $J$ along the quantization axis, while $\alpha_s(\omega)$ and $\alpha_t(\omega)$ are the scalar and tensorial part of the dynamical polarizability, respectively. In this expression, the tensorial contribution to the total polarizability has a dependence on the angle $\theta$ between the quantization axis and the laser electric field polarization. In our case, the Dy condensate is spin-polarized in its Zeeman ground state through the application of a magnetic field along $z$. Therefore we have that $\abs{M_J}=J=8$ and Eq.~\ref{eq:firstalphaofomega} can be written in the simplified form:
\begin{equation} \label{eq:totalpolar}
     \alpha(\omega)=\alpha_s(\omega)+\frac{3\cos^2\theta-1}{2}\alpha_t(\omega)\,.
\end{equation}

Both $\alpha_s(\omega)$ and $\alpha_t(\omega)$ depend on the frequency of the laser field and on the ground state angular momentum $J$. Their full theoretical expressions are given by:

\begin{equation} \label{eq:definizionedialphasealphat}
    \begin{split}
        &\alpha_s(\omega)=-\frac{1}{\sqrt{3(2J+1)}}\mathcal{F}_J^{(0)}(\omega)\\
        &\alpha_t(\omega)=\frac{2\sqrt{5J(2J-1)}}{\sqrt{6(2J+3)(J+1)(2J+1)}}\mathcal{F}_J^{(2)}(\omega)
    \end{split}
\end{equation}

where we have defined

\begin{equation} \label{eq:Fk}
    \begin{split}
        &\mathcal{F}_J^{(k)}(\omega)=\sum_{\beta'J'}(-1)^{J+J'}\begin{Bmatrix}
            1&1&k\\J&J&J'
        \end{Bmatrix}\abs{\mel{\beta J}{|\bm{d}|}{\beta' J'}}^2\cross\\
        &\frac{1}{\hbar}\left[\frac{1}{\omega_{\beta'J'}-\omega+\text{i}\gamma_{\beta'J'}/2}+\frac{(-1)^k}{\omega_{\beta' J'}+\omega+\text{i}\gamma_{\beta' J'}/2}\right]\,.
    \end{split}
\end{equation}

In this expression, the curly brackets represent a Wigner $6-j$ symbol, and $\mel{\beta J}{|\bm{d}|}{\beta' J'}$ is the reduced dipole matrix element evaluated between the atomic ground state $\ket{\beta J}$ and a generic excited state $\ket{\beta' J'}$ characterized by an energy $\hbar\omega_{\beta' J'}$ and natural linewidth $\gamma_{\beta' J'}$. The sum in Eq.~\ref{eq:Fk} is taken over all dipole-allowed transitions. 

In this paper we have experimentally investigated the region close to the $4f^{10}$$6s^2$ ($^5I_8$)$\rightarrow$$4f^{10}$($^5$$I_8$)$6s6p$($^1$$P_1$)$(8,1)_7$ resonance around 405~nm. Because of this, in the sum defined in Eq.~\ref{eq:Fk} this transition will give the most important contribution. 
From Eq.~\ref{eq:definizionedialphasealphat}, notice that one (two)-electron atoms, having $J=1/2\;(J=0)$ in their ground state, have no tensorial contribution to their ground state polarizability.

\subsubsection{\label{sec:level3}Polarizability of Dy in the 400-nm region}
\begin{figure}[!hbt]
\includegraphics[width=1\columnwidth]{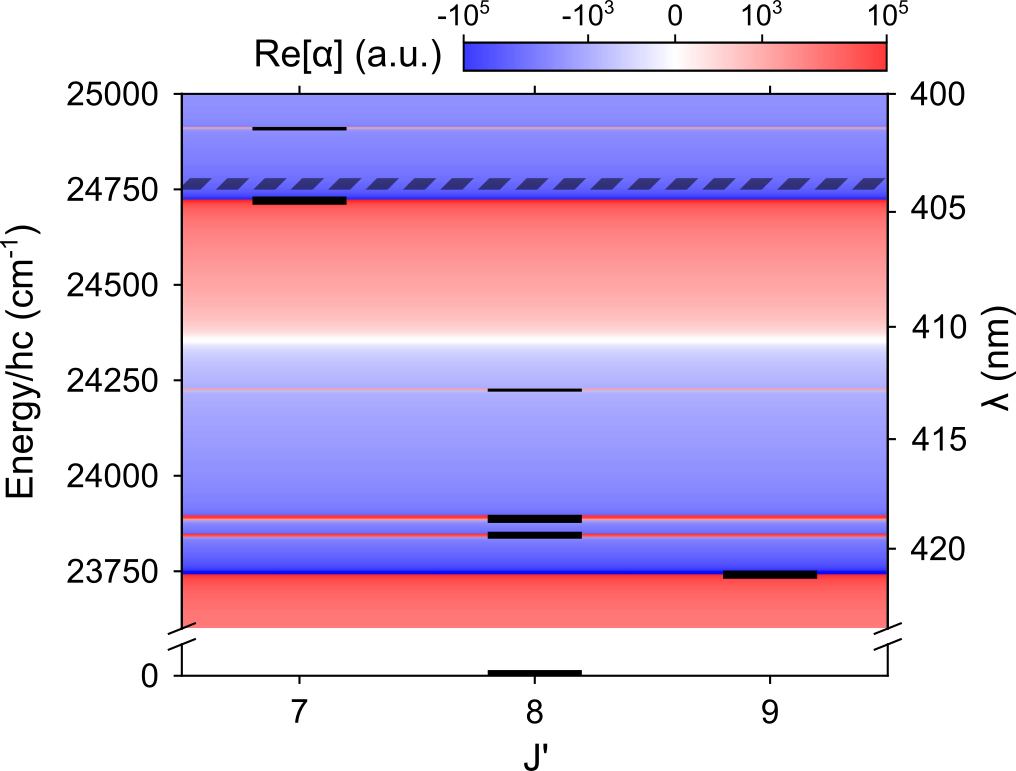}
\caption{\label{fig:levelscheme}Level scheme of Dy in the region of interest. Thicker black lines correspond to levels with larger linewidth. The background color scheme indicates the theoretical value of $\Re[\alpha_s-\alpha_t/2]$, which is the value of the total polarizability shown in Eq.~\ref{eq:totalpolar} for $\theta=\pi/2$. In the figure, blue (red) regions represent an overall repulsive (attractive) character. The dashed area represents the spectral region experimentally investigated in this work.}
\end{figure}

For our repulsive potential we decided to use light on the blue side of the 404.7~nm line of dysprosium, which is the transition from the ground state $4f^{10}$$6s^2$($^5$$I_8$) to the excited state $4f^{10}$($^5$$I_8$)$6s6p$($^1$$P_1$)$(8,1)_7$, see Fig.~\ref{fig:levelscheme} or Table~\ref{tab:levelscheme} for reference. It is a strong transition, having a linewidth $\gamma$ around $2\pi\times 30$~MHz \cite{Wi00}. We chose this region because laser diodes with output power of the order of 100~mW around 405~nm are commercially available.

\begin{table}[h]
\setlength{\tabcolsep}{6pt} 
\renewcommand{\arraystretch}{1.5} 
    \centering
    \begin{tabular}{c c c c }
    \hline
        level & $J'$ & $\lambda$(nm) & $\gamma/2\pi$(MHz) \\
        \hline
        $4f^{10}6s^2(^5I_8)$ & 8 & -- & --\\
        $4f^{10}(^5I_8)6s6p(^1P_1^o)(8,1)_9^o$ & 9 & 421.3 & 33.1 \\
        $4f^9(^6H^o)5d^2(^3F)(^8K^o)6s$ & 8 & 419.6 & 14.0 \\
        $4f^{10}(^5I_8)6s6p(^1P_1^o)(8,1)_8^o$ & 8 & 418.8 & 20.1 \\
        $4f^9(^6H^o)5d^2(^3F)(^8I^o)6s$ & 8 & 413.2 & 0.3 \\
        $4f^{10}(^5I_8)6s6p(^1P_1^o)(8,1)^o_7$ & 7 & 404.7 & 30.6 \\
        $4f^{10}(^5I_6)6s6p(^3P_2^o)(6,2)_7^o$ & 7 & 401.5 & 0.5\\
        \hline
    \end{tabular}
    \caption{Energy levels of dysprosium in the region of interest \cite{sansonetti2005handbook,Wi00}. Vacuum transition wavelengths from the ground state and linewidths are approximated at the first decimal position.}
    \label{tab:levelscheme}
\end{table}

\begin{figure}[!hbt]
\includegraphics[width=1\columnwidth]{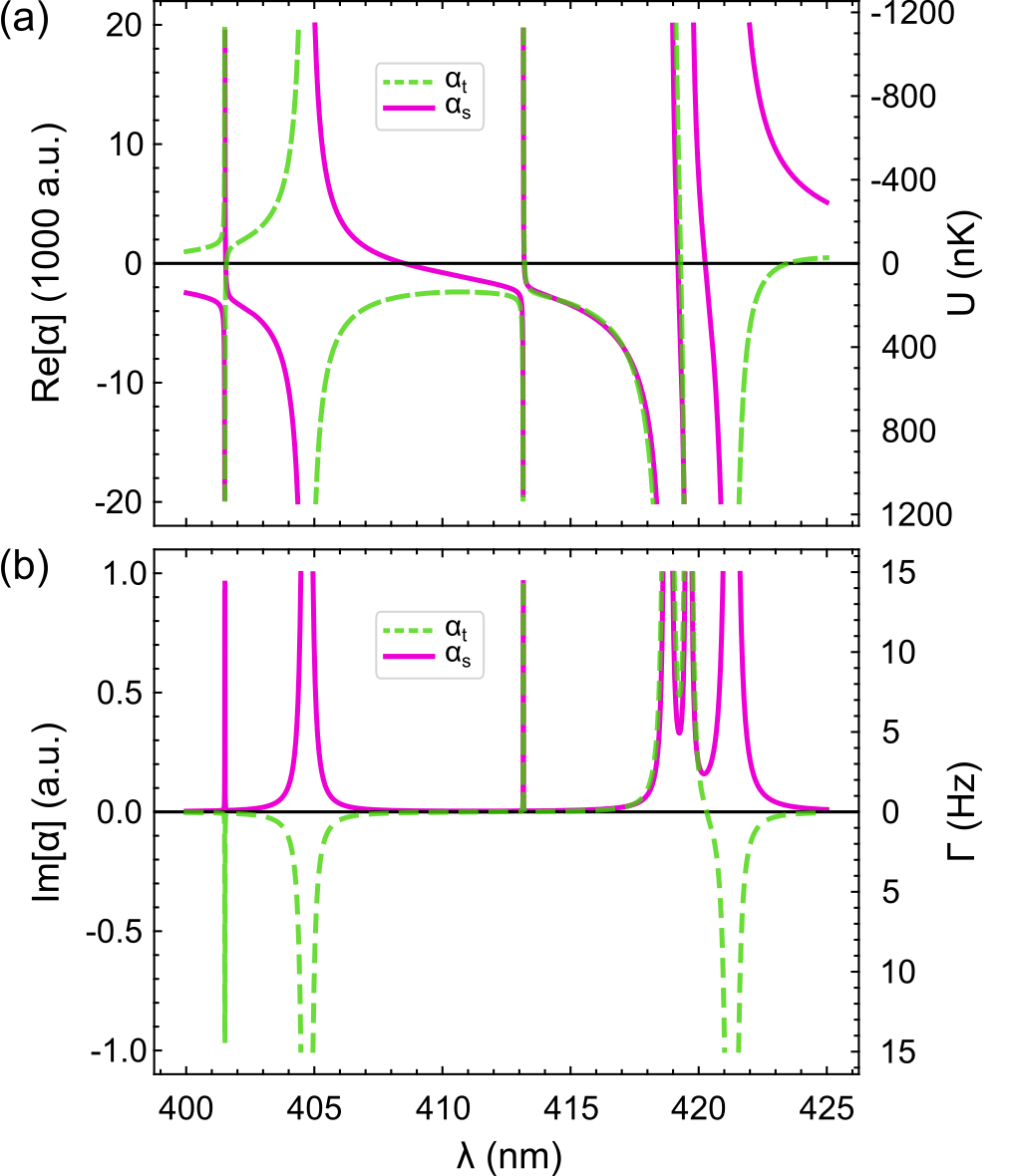}
\caption{\label{fig:predicted}  Theoretical polarizability of dysprosium in the blue region. Real (a) and imaginary (b) part of the scalar polarizability (continuous pink) and tensorial polarizability (dashed green). The potential strength $U$ and scattering rate $\Gamma$ is calculated considering a laser power of 1~mW on a Gaussian beam with a waist of $50\;\si{\micro\meter}$.}
\end{figure}

In Figure \ref{fig:predicted} we show the calculations of the scalar and tensorial parts of the polarizability in a range of wavelengths in the blue side of the spectrum. These plots are made using Eq.~\ref{eq:definizionedialphasealphat}, where the values for $\omega_{\beta'J'}$ and $\gamma_{\beta'J'}$ of the various transitions were taken from the NIST database \cite{sansonetti2005handbook,Wi00}. In calculating $\mathcal{F}_J^{(k)}$, it is useful to convert the reduced dipole matrix elements into linewidths using the formula:
\begin{equation}
    \abs{\mel{\beta J}{|\bm{d}|}{\beta' J'}}^2=(2J'+1)\frac{3\pi\epsilon_0\hbar c^3}{\omega_{\beta'J'}^3}\gamma_{\beta'J'}
\end{equation}
Looking at Fig.~\ref{fig:predicted}, we observe that the best spectral regions where to realize repulsive potentials are two, the first one in the approximate range (402-404)~nm, and the second one in the range (415-418)~nm. Both regions benefit from being on the blue side of a strong absorption line, but they differ in the sign of $\alpha_t$. From 402 to 404~nm, which is the region explored experimentally in this work, $\alpha_t$ is positive, meaning that the maximally repulsive polarizability is achieved setting $\theta=\pi/2$ in Eq.~\ref{eq:totalpolar}, and is given by $\alpha_s-\alpha_t/2$. This region is then best suited to realize an optical potential propagating along the quantization axis, since in this case $\theta$ is geometrically fixed at $\pi/2$. In the second region, $\alpha_s$ and $\alpha_t$ share the same sign. This means that the maximally repulsive polarizability is now given by $\alpha_s+\alpha_t$, obtained by putting $\theta=0$ in Eq.~\ref{eq:totalpolar}. This makes this region the optimal one to realize a repulsive potential with light that has a polarization pointing along the quantization axis, i.e. propagating perpendicular to the quantization axis.

\section{\label{sec:level2}Experimental setup}

\subsubsection{\label{sec:level3} Spectrally-filtered diode laser}

\begin{figure}[!hbt]
\includegraphics[width=1\columnwidth]{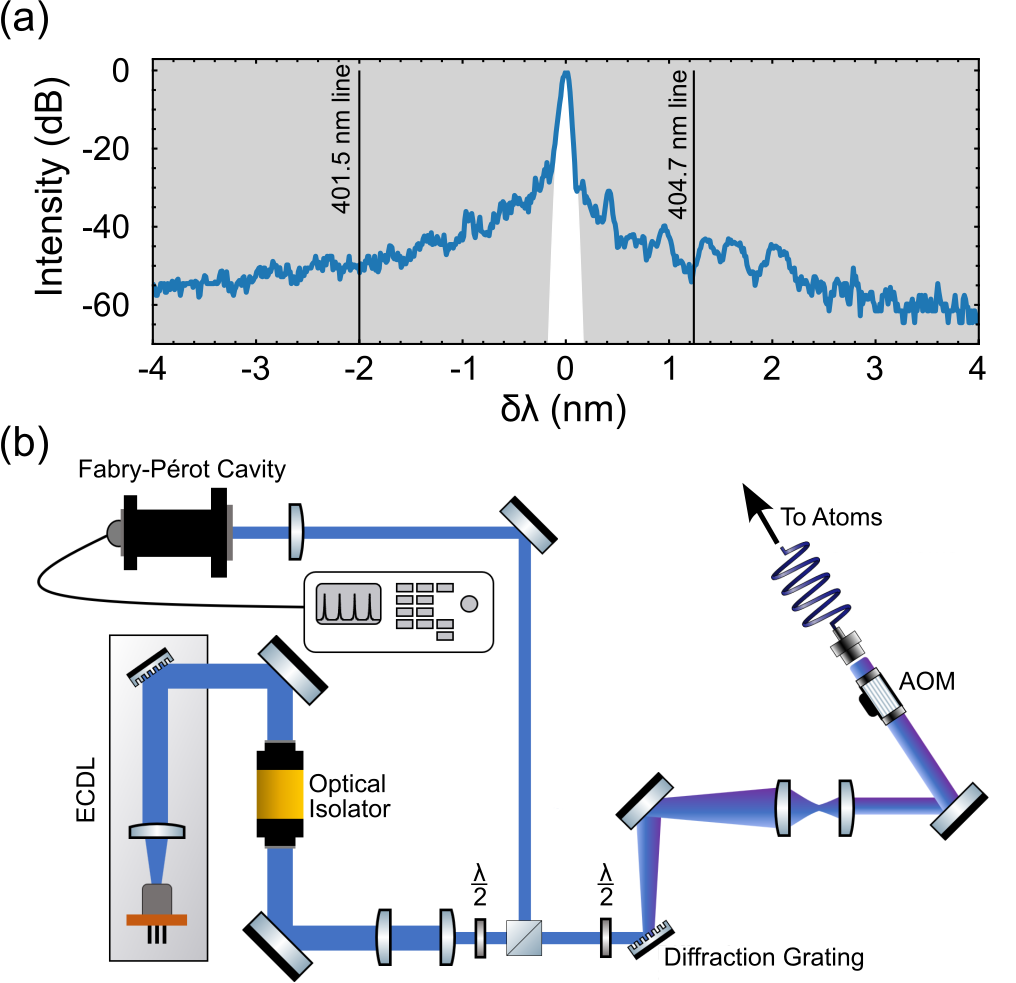}
\caption{\label{fig:expset}a) Spectrum of the laser diode in single-mode operation at a wavelength of 403.5 nm, captured on a CCD camera collecting the light coming from a diffraction grating. The long tails of the spectrum, dominated by the ASE (on top of the strongly suppressed secondary modes), cross both the 401.5 nm and the strong 404.7 nm resonances. The white region in the plot indicates the acceptance of the optical fiber used for filtering. b) Sketch of the experimental setup for the extended-cavity diode laser. The second diffraction grating, in combination with the single-mode optical fiber, filters out most of the amplified spontaneous emission of the laser.}
\end{figure}

The light source used in the experiment is a Nichia NDV4313 diode laser with a maximum output power of 120~mW near 405~nm. Under free-running conditions, its spectrum exceeds 1 nm in width and overlaps with the strong transition at 404.7 nm, making the laser unsuitable for atom trapping.
Mounted on an external cavity, realized with a holographic grating with 3600~grooves/mm in Littrow configuration, the laser operates in single-mode with 30 mW of output power and a tunability of about 1~nm changing the grating angle.
We monitor the laser spectrum with a scanning Fabry-Perot interferometer (Thorlabs SA200-3B), and we measure the laser wavelength with a wavelength meter having an accuracy of 600~MHz (HighFinesse WS-6).
A sketch of the laser setup and its typical spectrum in single-mode operation are shown in Figure~\ref{fig:expset}.\\

\begin{figure}[!hbt]
\includegraphics[width=1\columnwidth]{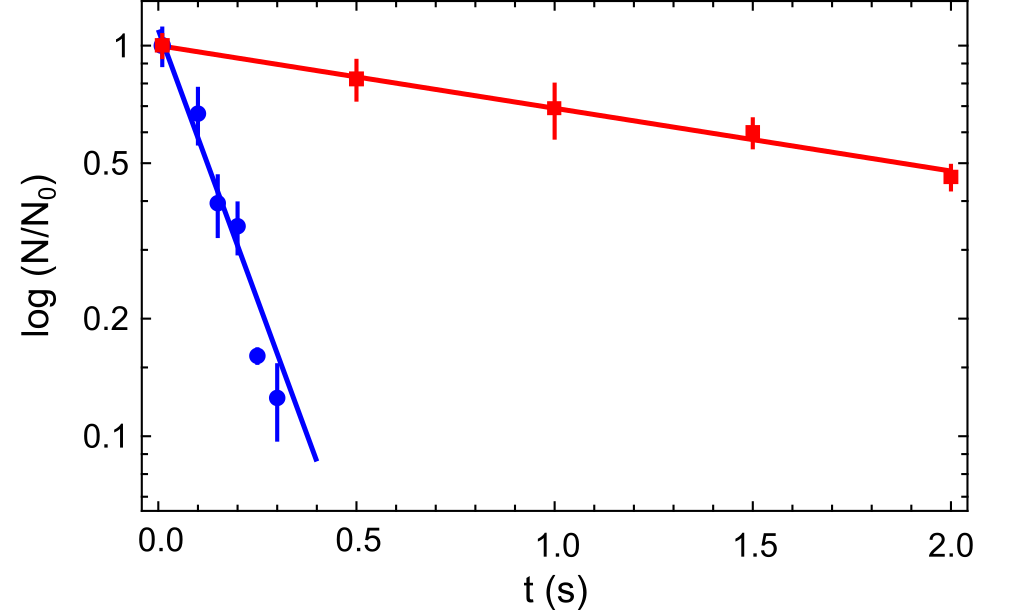}
\caption{\label{fig:lifetime} Measured lifetime of a dysprosium Bose-Einstein condensate irradiated by the blue light. The lifetime with additional spectral filtering after the extended cavity (red squares) is $(2.5\pm 0.3)$~s, much longer than the one without (blue dots), which is $(120\pm30)$~ms. Lines are exponential fits. The laser beam was Gaussian, with an average waist of 48 $\si{\micro\meter}$ (see text), a power of 2~mW, and a wavelength of 403.6~nm.}
\end{figure}

Besides the sharp peak at the central wavelength, the spectrum has a broad background of amplified spontaneous emission (ASE) (see Fig.~\ref{fig:expset} a), a fraction of which is in resonance with the atoms.
Indeed, we observe that the lifetime of the trapped atoms, see Figure~\ref{fig:lifetime} (blue dots), is significantly affected by the blue light. Even at low power and relatively large detuning (of about 1 nm), the resonant background considerably reduces the lifetime to hundreds of milliseconds.
To mitigate the resonant scattering
we further clean the laser spectrum, as displayed in Fig.~\ref{fig:expset} b).

The laser beam impinges onto a second diffraction grating, whose first diffraction order is injected into the fiber.
Due to the finite angular acceptance of the fiber, only a portion of the spectrum, the white region in Fig.~\ref{fig:expset}a, can be coupled. Considering the Mode Field Diameter (MFD) of the fiber, the spectral dispersion of the grating, and the coupling lenses, we estimate the bandwidth of the fiber to be around 0.11 nm. Note that this is of the same order of magnitude as the spectral resolution of our measurement given by the width of the central peak in Fig.~\ref{fig:expset}a.\\

The performance of this filtering technique can be evaluated by looking directly at the lifetime of the atoms, see Fig.~\ref{fig:lifetime}. Using filtered light (red squares), the lifetime significantly increases to several seconds, making our repulsive light usable in experiments. We will discuss in detail the lifetime in the next section.
Note that we can easily swap between filtered and unfiltered light by selectively coupling into the fiber either the zero or the first diffraction order of the grating.
A downside of this filtering technique is the loss of power due to the diffraction efficiency of the second grating, which is around 60\% for the first diffraction order.

As an alternative technique, it is possible to substitute the second grating in our scheme with a tilted clean-up filter. This method is more efficient in terms of power, since filters typically transmit up to around 95\% of the total power, but have a reduced performance in terms of filtering. For example, the Semrock filter model LL01-405 at the right angle cuts the resonant frequencies only up to three orders of magnitude, for the range of detunings of our experiment. One can of course improve the filtering by using two or more filters in series.

\subsubsection{\label{sec:level3} Atomic system}
The polarizability measurements are done on a Bose-Einstein condensate (BEC) of $^{162}$Dy atoms. A beam of Dy atoms is slowed and cooled by a Zeeman slower and a laser operating on the 421~nm transition line. The atoms are then collected in a magneto-optical trap (MOT) operating on the 626~nm line, transferred to an infrared (IR) optical dipole trap enhanced by an in-vacuum cavity, and finally trapped in two crossed IR beams where the final evaporation stage to condensation is realized. More details of the setup can be found in \cite{Lu18}.

The light near 405~nm, coming from the fiber, passes through a $\lambda/4$ waveplate in series to have a pure linear polarization, and is reflected by a polarizing beam splitter. Then it passes through another half waveplate and is reflected by a mirror and focused on the atoms by a 200~mm lens on the horizontal plane. The second half waveplate sets the angle between the laser polarization and the magnetization of the atoms, which is vertical. The final mirror has two tilting screws regulated by two step motors; we calibrated the tilting angle on both axes as a function of the pulse duration given to the step motors. The laser power $P$ is measured by two different calibrated power meter heads to minimize errors.

\section{\label{sec:level2} Measurements and discussion}

\subsubsection{\label{sec:level3} Measurement techniques}
\begin{figure}[!hbt]
\includegraphics[width=1\columnwidth]{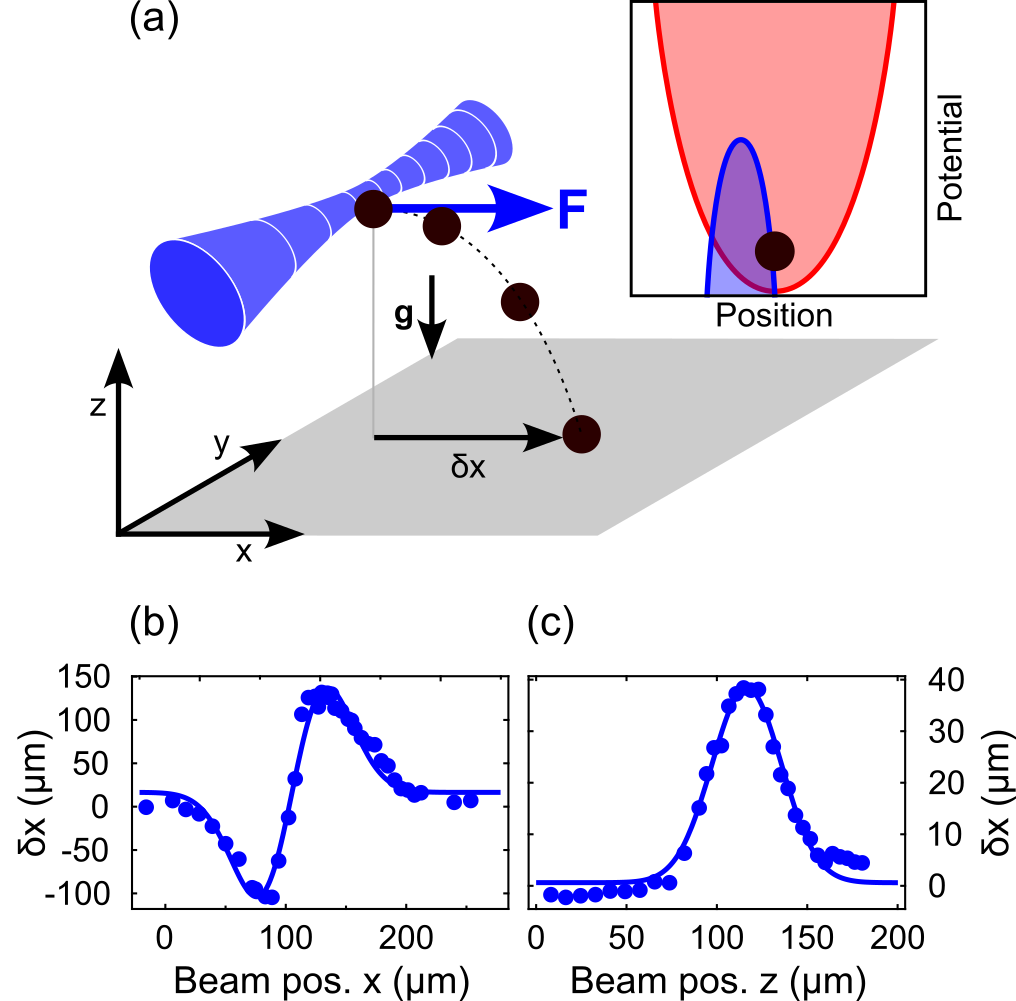}
\caption{\label{fig:sketch} a) Sketch of the method used to measure the polarizability. The repulsive blue beam is switched on for a short time $\delta t$ and imparts a kick to the atoms while they are confined in the attractive infrared trap. In the subsequent expansion in the absence of optical trapping, the momentum kick translates into a displacement $\delta x$ of the atomic center of mass in the horizontal plane $xy$, which is proportional to the dynamical polarizability. The $x$-axis is chosen to be the direction of the atomic displacement. b-c) Displacement of the atomic center-of-mass $\delta x$ on the horizontal plane as a function of the light beam position on the $x$-axis (b) and the $z$-axis (c).}
\end{figure}

Our measurement technique relies on the momentum imparted to the atomic cloud by a pulse of repulsive light, as sketched in Fig.~\ref{fig:sketch}. The pulse duration $\delta t$ is set by an acousto-optic modulator and a shutter before the fiber. After the light pulse, we let the atoms expand freely and we detect them after a time of flight $t_{exp}$ by absorption imaging along the vertical axis.

If we consider a Gaussian light beam propagating along $y$ with waists $w_x$ and $w_z$, and assume that the BEC is point-like, the atomic cloud will be subjected to a force in the direction $x$ perpendicular to the propagation one \footnote{A force along the $z$ direction is also applied, given by $-\dd U/\dd z$. However, we do not measure this component of the force because with our imaging we integrate in the $z$ direction and we observe only the $xy$ plane.},
\begin{equation}
    F(x, z, \omega)=-\frac{\dd U(x,z, \omega)}{\dd x}=\frac{\Re[\alpha(\omega)]}{2\epsilon_0 c}\frac{\dd I(x,z)}{\dd x}\,,
\end{equation}
with $I(x,z)$= $I_0 e^{-2x^2/w_x^2}$$e^{-2z^2/w_z^2}$, where $I_0$ is the maximum laser intensity. Immediately after the pulse, we let the system expand for a time $t_{exp}$ switching off the optical trapping and we measure the displacement $\delta x$ of the cloud along in the horizontal plane. The displacement is proportional to the force as $\delta x = \left(F \,t_{exp}\, \delta t \right)/m $. The force, hence $\delta x$, has a dispersive behaviour as a function of the beam position with a maximum at $x=w_x/2$ and $z=0$:
\begin{equation}
F_{max}(\omega)=\frac{2\sqrt{e}\Re[\alpha(\omega)]P}{\pi \epsilon_0 c \, w_z w_x^2}\,,
\end{equation}
where $P$ is the laser power. If we take into account the finite dimensions of the BEC and we average over a Thomas-Fermi distribution calculated for $20\times 10^3$ atoms in our IR harmonic trap, the maximum force decreases by 7$\%$. Measuring $\delta x$ for different beam positions in the $xz$ plane
allows us to derive the waist, as shown in Fig.~\ref{fig:sketch},
directly at the atom position, thus minimizing sources of errors typical in non-local measurements. Moreover, our measurement has a cubic dependence on the waist, while a measurement by the trap frequencies depends on the fourth power of the waist. Moving the beam along the horizontal axis, Fig.~\ref{fig:sketch}(a), we see that the displacement on the horizontal plane changes sign when we cross the atoms' position. The measured $\delta_x$ has the shape of the spatial derivative of the beam intensity. Instead, along the vertical direction, Fig.~\ref{fig:sketch}(b), we observe on the horizontal plane just the intensity profile, i.e. a simple Gaussian shape. The fitted values of the waists are $w_x$=(58$\pm 7)$ $\si{\micro\meter}$, $w_z$=(38$\pm 3)$~$\si{\micro\meter}$, where the errors come from both the fitting errors and the calibration uncertainties. 


The atoms undergo a maximum displacement $\delta x_{max}$,
\begin{equation} \label{eq:maxdisp}
\delta x_{max}(\omega)=\frac{F_{max} t_{exp} \delta t}{m}=\frac{2\sqrt{e}\Re[\alpha(\omega)]P \,t_{exp}\delta t}{\pi \epsilon_0 m c\,w_z w_x^2}\,.
\end{equation}

\begin{figure}[!hbt]
\includegraphics[width=1\columnwidth]{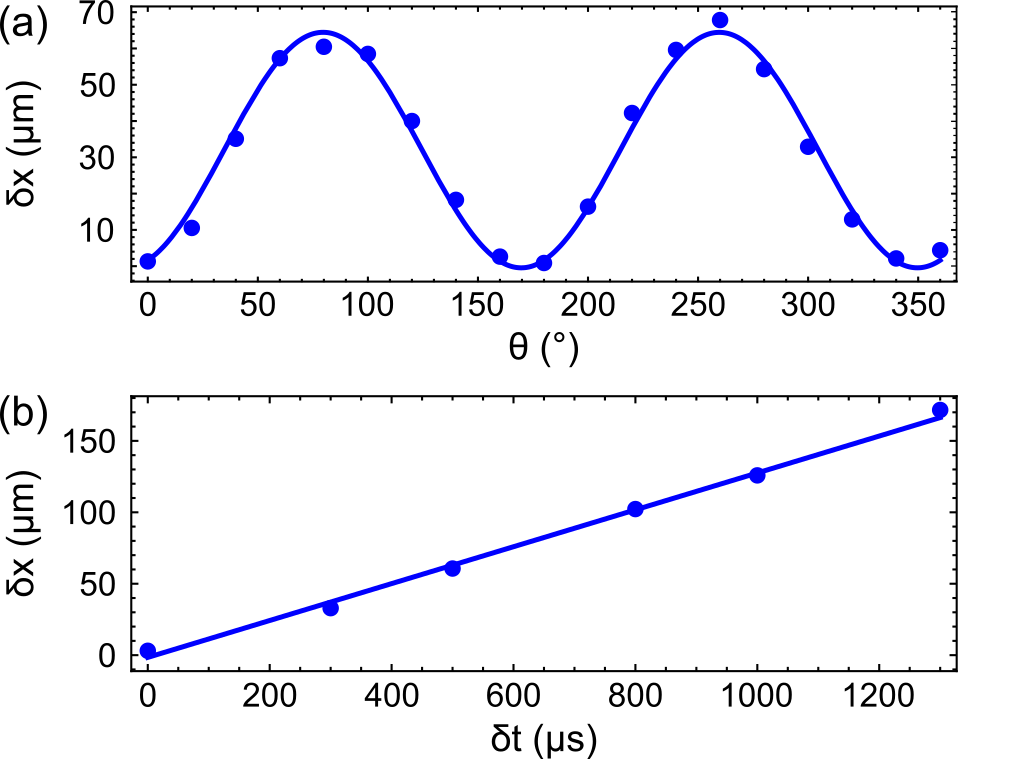}
\caption{\label{fig:displacement} Measurement of the scalar and tensorial polarizabilities: 
(a) Displacement $\delta x$ of the atomic center-of-mass as a function of the angle $\theta$ of the light polarization with respect to the magnetization axis, together with the sinusoidal fit by Eq.~3 (line). (b) Displacement $\delta x$ of the atomic center-of-mass as a function of the laser pulse duration $\delta t$ together with a linear fit (line). The experimental error bars are within the marker dimension.}
\end{figure}

Fixing the beam position to have $\delta x = \delta x_{max}$, a first measurement of the maximum displacement is done as a function of the polarization angle with respect to the magnetization, to find the relative contribution of the scalar and tensorial polarizabilities. The results of this are shown in Fig.~\ref{fig:displacement}(a), together with a fit done using Eq.~3. 

The fit parameters give the relative contributions of the scalar and tensorial polarizabilities. At this point we fix the half waveplate at the angle for which the atomic displacement is maximum, that is when the light polarization is perpendicular to the atomic magnetization, and we instead change the laser power or the laser pulse duration. Doing this we obtain a curve like that in Fig.~\ref{fig:displacement}b, if we remain in a regime where the displacement is linear. From the linear fit coefficient we can derive the absolute polarizability, inverting Eq.~\ref{eq:maxdisp}.

\subsubsection{\label{sec:level3} Experimental polarizability}
\begin{figure}[!hbt]
\includegraphics[width=1\columnwidth]{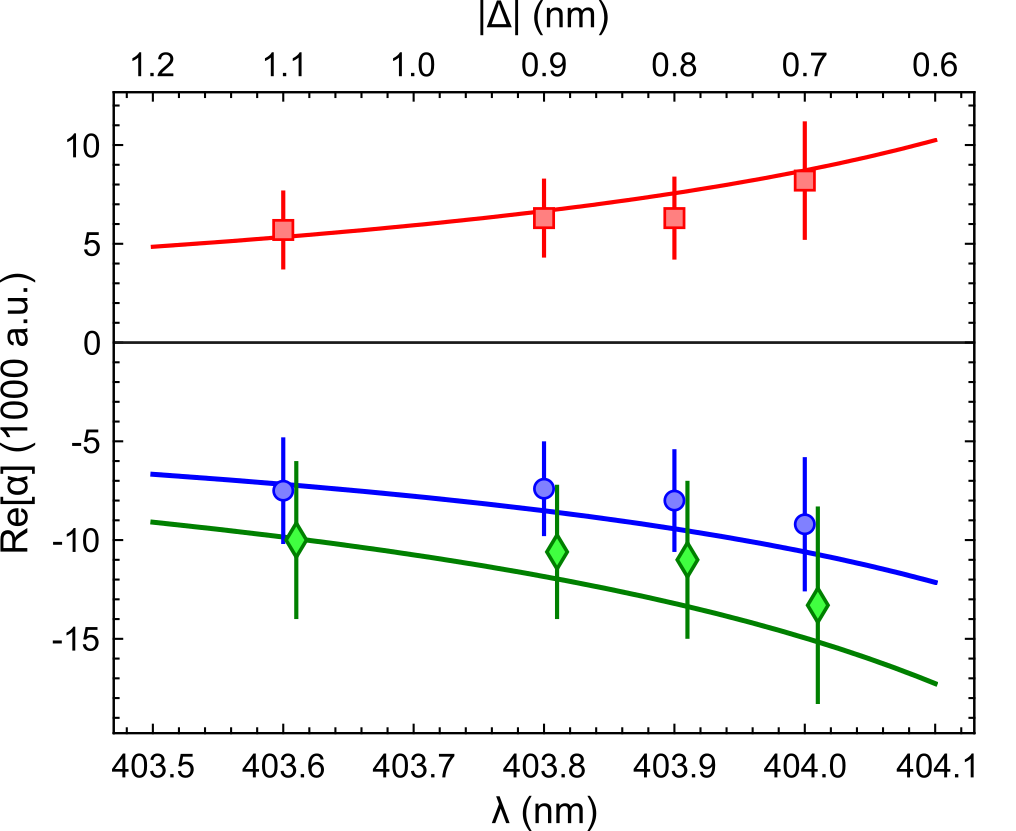}
\caption{\label{fig:respol} Experimental polarizabilities as a function of the wavelength $\lambda$ vs theoretical predictions. Scalar polarizability $\alpha_s$ (blue dots), tensorial polarizability $\alpha_t$ (red squares), and maximum polarizability $\alpha_s - \alpha_t/2$ (green diamonds) show a good agreement with the theoretical predictions (lines). The green dots are shifted by 0.01~nm to the right for clarity. The upper horizontal axis indicates the detuning $\Delta$ from the transition at 404.7 nm} 
\end{figure}

We have repeated the polarizability measurements for a few wavelengths on the blue side of the 404.7~nm line. The measurements are shown in Fig.~\ref{fig:respol}, together with the theoretical predictions from Eq.4. Also, the maximum polarizability is shown, occurring when the light polarization is perpendicular to the atomic magnetization. Considering the error bars, mainly due to the waist measurement, both the scalar and tensorial polarizabilities are in good agreement with the theoretical expectations. We did not extend the study closer to resonance, as the lifetime due to photon scattering becomes more critical. The real part of the polarizability scales with detuning $\Delta$ as $\Delta^{-1}$, while the photon scattering rate, proportional to the imaginary part of the polarizability, as $\Delta^{-2}$  \cite{Gr00}.\\

\subsubsection{\label{sec:level3} Lifetime measurements}
Concerning the lifetime of the BEC in the presence of blue light, the interpretation of what we observe in Fig.~\ref{fig:lifetime} must take into account the repulsive effect of the optical potential. Under the conditions of Fig.~\ref{fig:lifetime}, i.e. laser power equal to 2~mW and wavelength of 403.6~nm, we expect a polarizability of about -10000~a.u. and a lifetime of about 400~ms if the atoms stay in the high-intensity region of the blue laser. However, the repulsive potential tends to expel the atoms from the high-intensity region, leading to a new equilibrium position inside the combined IR and blue traps, where the intensity of the blue laser is lower. We have modeled the combined traps and found that the equilibrium position moves by about one waist along $x$, meaning that the atoms feel a blue intensity reduced by a factor $e^{-2}$. Correspondingly, the lifetime increases by $e^{2}$, which means up to 3~s, of the same order of magnitude as the experimental value, 2.5~s. We note that this value is of the same order as the lifetime of the BEC in the infrared trap alone. At the current stage, a precise measurement of the effect of the blue light on the lifetime is therefore not possible. Further experiments in which the repulsive potential gives a dominant and well-controlled contribution to the shape of the optical trap, for example in a box-like configuration, are required.\\

\section{\label{sec:level2} Conclusions}

In conclusion, we reported the realization of a repulsive dipole potential for dysprosium atoms using a blue laser. We measured the scalar and tensor polarizabilities near the 404.7~nm line and found good agreement with the theory. At the wavelength $\lambda=404$ nm where the scalar and tensor polarizabilities are of the same order of magnitude, we observe a very large maximum value of about -13000~a.u. for the total polarizability. This means that we can realize a repulsive potential of depth $U=k_B\times$500~nK with only 1~mW of blue light on a Gaussian beam with a relatively large waist of 60~$\si{\micro\meter}$. The lifetimes are of the order of 1~s, and are improved by the fact that the atoms do not stay in the highest intensity regions. 
This type of optical potential can be used for a variety of experiments on dipolar superfluids and supersolids, as discussed in the introduction.\\

Since the tensor contribution to the polarizability depends on the total angular momentum projection $M_J$, the large values of $\alpha_t$ observed in this work can also be exploited to create controllable state-dependent potentials for dysprosium atoms in different Zeeman states, such as bilayer potentials \cite{Du23}. By controlling the strength of the tensor polarizability with the polarization angle $\theta$, it might be possible to construct different types of state-dependent optical lattices. For example, in the wavelength range 407-410 nm where $\alpha_t>\alpha_s$, one could realize anti-magic optical lattices with opposite polarizability for two different atomic states  \cite{PhysRevLett.132.153401}, thus realizing extremely short-wavelength optical lattices with an effective lattice spacing of $\lambda/4\approx$ 100 nm. In this case, the lattice spacing would be comparable to the dipolar length, which determines the characteristic distance of long-range dipole-dipole interactions. This would open the way to study long-range interacting lattice models with Dy-based atomic quantum simulators.


\begin{acknowledgments}
Funded by the European Union (ERC, SUPERSOLIDS, n.101055319) and by the QuantERA Programme, project MAQS, under Grant Agreement n.101017733, with funding organisation Consiglio Nazionale delle Ricerche. We acknowledge support from the European Union - NextGenerationEU for the “Integrated Infrastructure initiative in Photonics and Quantum Sciences” - I-PHOQS [IR0000016, ID D2B8D520, CUP B53C22001750006] and for the PNRR MUR project PE0000023-NQSTI . Technical assistance from A. Barbini, F. Pardini, M. Tagliaferri and M. Voliani is gratefully acknowledged.
\end{acknowledgments}

\bibliography{\bibname}
\end{document}